# Importance of the spectral emissivity measurements at working temperature to determine the efficiency of a solar selective coating

T. Echániz [a], I. Setién-Fernández [a], R.B. Pérez-Sáez [a,n], C. Prieto [b], R. Escobar Galindo [b,1],

M.J. Tello [a]

[a] Departamento de Física de la Materia Condensada, Facultad de Ciencia y Tecnología, Universidad del País Vasco, Barrio Sarriena s/n, 48940 Leioa,

Bizkaia, Spain

[b] Instituto de Ciencia de Materiales de Madrid, (ICMM-CSIC), Campus Cantoblanco, 28049 Madrid, Spain

Abstract

The total emissivity of the absorbing surfaces is a critical parameter in the calculation of the radiative thermal losses in solar thermal collectors. This is because the radiative heat losses have a significant economic impact on the final cost of the electricity produced in a solar thermal plant. This paper demonstrates the need to calculate the total emissivity from spectral emissivity measurements at the working temperature of the solar thermal collector, instead of using extrapolated values from spectral emissivities measured at room temperature. Usual uncertainties produced by the estimation of the total emissivity, in which its temperature dependence is only introduced by the Planck function, are analyzed.

1. Introduction

Solar thermal collectors (STC) are especially relevant in solar thermal devices for producing heat from the sunlight (T < 150 °C), or producing electricity in concentrated solar power (CSP) plants (250 < T < 800 °C). A critical component of solar thermal collectors is the surface where the conversion of solar radiation into useful heat occurs. These solar absorbing surfaces (SAS) require very high solar absorptivity $\alpha(T)$ in visible and near infrared (NIR) wave- lengths and low total emissivity $\varepsilon T(T)$ in the NIR–mid infrared (MIR) spectral region [1]. These physical requirements must be achieved at the

SAS working temperature, which can be around 600 °C in common CSP plants. There is a wide literature, including several reviews, see for example Refs. [1–5], studying SAS. At this point it is important to outline some of the early works on this subject where the basis of the main properties of the selective coatings was set [6–10]. However, the analysis of the SAS characterization literature demonstrates that, in most cases, the total emissivity is obtained by making use of two approximate methods. Both of them make use of experimental data between room temperature and 100 °C, and assume that the spectral emissivity $\epsilon_\lambda(\lambda,T)$ of the SAS is constant with temperature. In one of these methods the total emissivity is measured at a fixed temperature, usually at T < 100 °C using a commercial emissometer. In the other method, the $\epsilon T(T)$ is obtained by using the reflectivity spectrum measured at room temperature, R($\lambda$), and extrapolating to working temperature T using the following integral:

$$\epsilon_T(T) = \frac{\int_0^\infty [1 - R(\lambda)] L(\lambda, T) d\lambda}{\int_0^\infty L(\lambda, T) d\lambda} \tag{1}$$

where L($\lambda$,T) is the Planck function for the emission of a blackbody. Both methods, the emissometer and reflectivity one, can give $\epsilon T(T)$ values that significantly differ from the real sample emissivity at working temperature because the coating spectral emissivity has usually a temperature dependence in the range of wavelengths where radiative transfer occurs. In addition, as it can be observed in other materials [11] the coating can even show an atypical temperature spectral emissivity behavior. Therefore, Eq. (1) should be replaced by

$$\epsilon_T(T) = \frac{\int_0^\infty \epsilon_\lambda(\lambda, T) L(, \lambda T) d\lambda}{\int_0^\infty L(\lambda, T) d\lambda} \tag{2}$$

which requires the measurement of the spectral emissivity $\epsilon_\lambda(\lambda,T)$ at the working temperature [12].

A simple calculation shows that small differences between the $\epsilon_T(T)$ values calculated by using Eqs. (1) and (2) may be important in defining the efficiency of a solar selective coating. An effective coating requires $\alpha(T) > 0.95$ and $\epsilon_T(T) < 0.05$ at working temperature (i.e. 600 °C). Therefore, certain difference between the values of the total emissivity obtained by Eqs. (1) and (2) produces an equivalent difference in the radiative thermal losses. These energy losses in the SAS increase proportionally to $T^4$ [4]. As a consequence, the differences between using Eqs. (1) or (2) may have significant economic impact on the final cost of the electricity produced in a solar thermal plant.

This paper is focused on demonstrating, from an experimental point of view, that the analysis of the efficiency of a solar coating at its working temperature requires the calculation of $\epsilon_T(T)$ from emissivity spectrum measured at that temperature. A stack with two cermet layers of silicon nitride with different amounts of molybdenum over a silver infrared mirror layer is used as solar coating. The experimental measurements were carried out between 250 and 600 °C and the experimental results obtained are compared with those calculated by the usual approximate methods.

2. Experimental results

Fig. 1 shows a schematic representation of the solar selective coating used in this paper. Experimental details about the coating preparation as well as the physical and chemical characterization have been given elsewhere. See Ref. [13] and references cited therein.

Spectral emissivity measurements were carried out with a homemade IR-radiometer [14] at moderate vacuum (E $10^{-3}$ mbar) during five heating cycles between room temperature and 600 °C. This moderate vacuum is used in a large number of industrial applications of solar collectors. Normal spectral emissivity is measured in each heating cycle at 15 different temperatures following the sequence shown in Ref. [12].

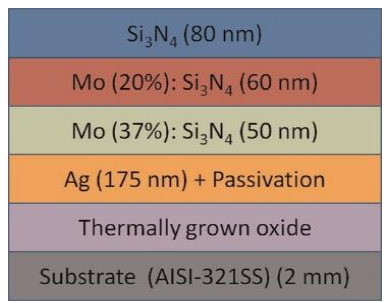

Fig. 1. Schematic representation of the solar selective coating.

The radiometer allows an accurate detection of the spectral thermal radiation as well as its fast processing. The sample holder permits directional measurements and the sample chamber ensures a controlled atmosphere and moderate vacuum. The radiometer calibration was carried out using a modified two-temperature method [15] and the emissivity spectra were obtained applying the blacksur method described in Ref. [16]. The maximum combined standard uncertainty varies between 1% and 10% depending on wavelength and temperature [17]. The average uncertainty value is around 4%. The sample temperature was measured by means of two spot-welded thermocouples on the sample surface.

In Fig. 2 we plot the emissivity versus the heating cycle number for two temperatures and four wavelengths. Similar plots were found at other temperatures and wavelengths. It should be noted that after the third heating cycle changes in the emissivity value are smaller than the uncertainty in its measurement. It can be stated that thermal stability of the coating emissivity is ensured after a few initial cycles. It is an essential property if one wants to use a coating in a STC.

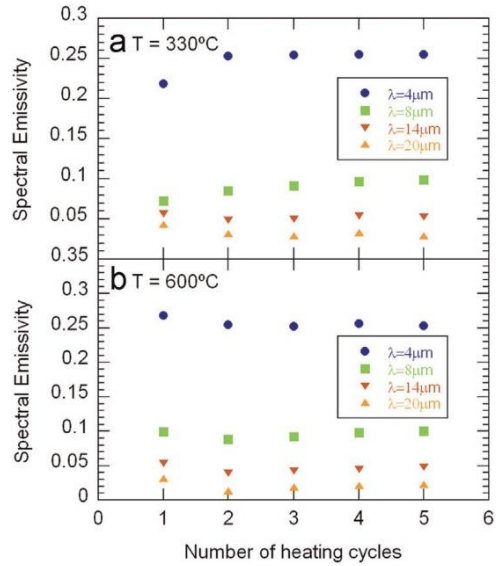

Fig. 2. Coating emissivity ε(λ,T) as a function of heating cycle for 4, 8, 14 and 20 μm at (a) T = 330 °C and (b) T = 600 °C.

In Fig. 3, the coating normal spectral emissivity at eight temperatures, between 250 and 600 °C, during the fifth heating cycle is plotted. Temperature dependence can hardly be detected and therefore a closer look has to be taken. In Fig. 4 the emissivity dependence on temperature at four different wavelengths is shown. Even though the emissivity usually increases with temperature [18,19], at short wavelengths such as 2 mm the emissivity decreases, whereas at longer wavelengths little or no evolution is appreciated. This behavior is crucial, since even small variations may be sufficient to give appreciable difference between calculated total emissivities from Eqs. (1) and (2). It is precisely the value of $\varepsilon_T(T)$ which indicates if a coating shows good performance for high temperature solar harvesting.

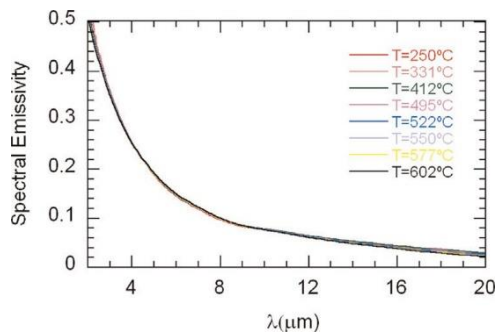

Fig. 3. Normal spectral emissivity ε(λ,T) of the selective coating for eight different temperatures measured during the fifth heating cycle.

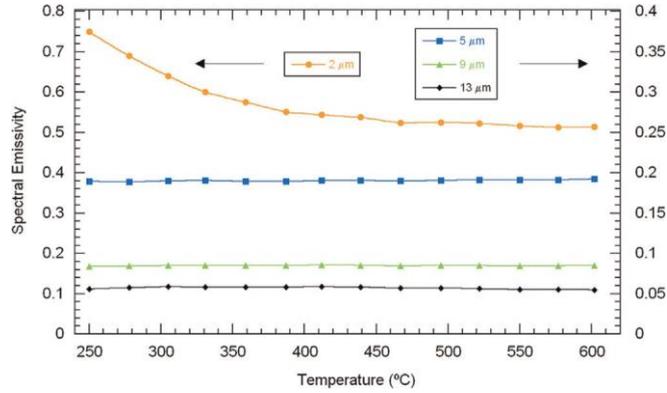

Fig. 4. Emissivity temperature dependence at 2, 5, 9 and 13 μm.

The comparison between the coating and substrate spectral emissivity values is an interesting feature when characterizing the coating efficiency. Fig. 5 shows the normal spectral emissivity of the coating and the substrate at 600 °C. It is observed that the coating spectral emissivity is appreciably lower than that of the substrate only for λ > 5 μm.

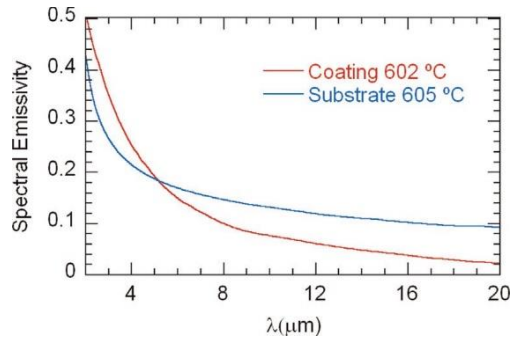

Fig. 5. Normal spectral emissivity ε(λ,T) of the selective coating and the substrate at working temperature.

The interest of the results is that they can be used to calculate and analyze one of the fundamental physical parameters in a thermal solar coating: the radiation energy losses at the working temperature. For this calculation the knowledge of $\varepsilon_T(T)$ at this temperature is necessary. What follows the differences that may exist between the calculation performed by Eq. (1) and those made with Eq. (2) is analyzed for the coating shown in Fig. 1. Here it should be noted that the integral calculation in Eqs. (1) and (2) is only accurate in the range of wavelengths where the emissivity spectra is measured with the radiometer ($\lambda_1 \le \lambda \le \lambda_2$) and, it requires an approximation out of that range, that is in the ranges $0–\lambda_1$ and $\lambda_2–\infty$. However, if we now consider that most of the losses occur in the range ($\lambda_1 \le \lambda \le \lambda_2$), the approximation is excellent [12].

In Fig. 6 the total emissivity ($\varepsilon_T(T)$) of the coating and the substrate as a function of temperature are plotted. In the figure we show the values of $\varepsilon_T(T)$ for the coating obtained from Eq. (1) using the room temperature reflectance spectrum [4], as well as those calculated by the Eq. (2) using the emissivity spectrum depending on the temperature. The differences between the values of $\varepsilon_T(T)$ for this coating at 400 °C and 600 °C are around 8%. Furthermore, significant differences in the temperature dependence of the $\varepsilon_T(T)$ curves obtained from Eqs. (1) and (2) are observed. The different temperature dependence of $\varepsilon_T(T)$ makes that the selectivity ratio $\xi = \alpha/\varepsilon_T$ also varies with T. Thus, the $\xi$ value at 400 °C is better with the $\varepsilon_T(T)$ calculated using Eq. (1) and, however, the one calculated with the $\varepsilon_T(T)$ calculated using Eq. (2) becomes better at 600 °C. For this temperature $\xi_{600 °C} = 4.1$ for Eq. (2) and 3.7 for Eq. (1) [13]. Therefore, for the coating studied in this work, the difference between the values of radiative thermal losses calculated with Eqs. (1) and (2) may vary by about 8%. Obviously, these differences increase if the dependence of the spectral emissivity with temperature increases. Additionally, one must take into account that in the case of the coating studied in this work, the spectral emissivity dependence on temperature shows an emissivity decrease at short wavelengths that produces attenuation on the temperature dependence of the total emissivity.

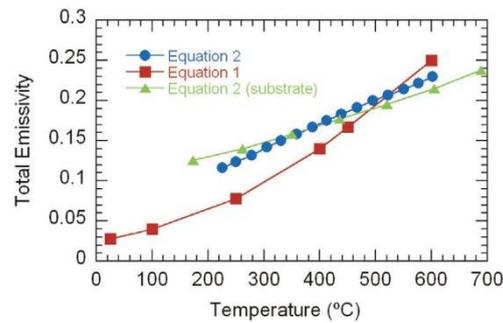

Fig. 6. Total normal emissivity $\varepsilon_T(T)$ of the coating obtained from radiometric measurements using Eq. (2), from reflectivity measurements using Eq. (1). The substrate results are also plotted.

Similar results, although with minor differences between the total emissivities calculated with Eqs. (1) and (2), were found in a coating with double cermet layer of silicon oxide with different amounts of molybdenum over a silver infrared mirror layer [12]. The close agreement between the radiometric and reflectivity measurements in Fig. 10 of Ref. [12] happens because in that case, due to the multilayer architecture, the temperature dependence of the spectral emissivity of the coating is weaker than the coating studied in this paper. Finally, the results of Fig. 5 also suggest that the total normal emissivity of the coating, and also the substrate, shows a quasi-linear behavior, in agreement with the electro- magnetic theory predictions [20].

3. Conclusion

In this paper, the total emissivity values of a selective absorber surface, in the working temperature range (250–600 °C), obtained from spectral emissivity measurements are compared with those obtained from room temperature reflectivity data. Both values will only agree when the spectral emissivity does not depend on temperature within the wavelength range in which the radiative transfer is made. The differences between the two values introduce significant discrepancies in the calculation of radiative thermal losses and therefore influence, significantly, in the calculation of the economic profitability of solar power plants. Therefore, the interest of measuring the temperature dependence of the spectral emissivity is fully justified. In addition, it is essential, for solar selective coatings applications to assess the emissivity changes with heating cycles and heating rate, which can only be proven with spectral measurements at different temperatures. Another important aspect for the applications that can be studied by means of radiometric methods is the detection of possible atypical temperature or wavelength dependences.

Finally, it has been shown that a solar coating qualified as efficient using Eq. (1) may, in fact, not be efficient when a more realistic calculation is done by introducing in Eq. (2) experimental data of the dependence of the spectral emissivity with temperature.

Acknowledgments

This work has been carried out with the financial support of the ETORTEK and SAIOTEK 2013 program (Project numbers S-PC08UN07 and S-PE13UN123 respectively) of the Basque Government in collaboration with CIC-Energigune. T. Echániz acknowledges the Basque Government their support through a Ph.D. fellowship.